\documentclass[11pt,tightenlines,showpacs,superscriptaddress]{revtex4}

\usepackage{amssymb,amsmath,graphicx}

\begin{document}

\title{Delayed feedback control of fractional-order chaotic systems}

\author{A. Gjurchinovski}
\email{agjurcin@pmf.ukim.mk}
\affiliation{Institute of Physics, Faculty of Natural Sciences and Mathematics, 
Saints Cyril and Methodius University, P. O. Box 162, 1000 Skopje, Macedonia}

\author{T. Sandev}
\email{trifce.sandev@avis.gov.mk}
\affiliation{Radiation Safety Directorate, Partizanski odredi 143,
P. O. Box 22, 1020 Skopje, Macedonia}

\author{V. Urumov}
\email{urumov@pmf.ukim.mk}
\affiliation{Institute of Physics, Faculty of Natural Sciences and Mathematics, 
Saints Cyril and Methodius University, P. O. Box 162, 1000 Skopje, Macedonia}

\date{20 August 2010}

\begin{abstract}
We study the possibility to stabilize unstable steady
states and unstable periodic orbits in chaotic 
fractional-order dynamical systems by the 
time-delayed feedback method. 
By performing a linear stability analysis, 
we establish the parameter ranges for successful 
stabilization of unstable equilibria in the plane 
parametrizad by the feedback gain and the time delay. 
An insight into the control mechanism 
is gained by analyzing the characteristic equation of the
controlled system, showing that the control 
scheme fails to control unstable equilibria having an 
odd number of positive real eigenvalues.
We demonstrate that the method can also stabilize 
unstable periodic orbits for a suitable choice of the
feedback gain, providing that the time delay is chosen
to coincide with the period of the target orbit.
In addition, it is shown numerically that delayed feedback control 
with a sinusoidally modulated time delay significantly 
enlarges the stability region of the 
steady states in comparison to the classical 
time-delayed feedback scheme with a constant delay.
\end{abstract}


\maketitle

\section{Introduction}

The fractional order dynamical systems have attracted remarkable
attention in the last decade. Many authors have studied the 
chaotic and hyperchaotic dynamics of various fractional order systems, such as
those of Duffing, Lorenz, R\"{o}ssler, Chua, L\"{u}, Chen, etc.,
which are introduced by changing the time derivative in the corresponding
ODE systems, usually with the fractional derivative
in the Caputo or Riemann-Liouville sense of order $0<\alpha<1$
\cite{Ge_Ou,Li_Liao_Yu,Li_Chen,Li_Peng,Zhang et al}. One interesting
problem is to analyze the lowest value of parameter $\alpha$ under
which fractional order dynamical systems show chaotic or
hyperchaotic behaviors. Stability analysis, synchronization and
control of fractional order systems by using different
techniques are also widely investigated
\cite{Matignon,Chen_Moore,Deng_Wu,Deng_Li,Deng1,Deng2,Deng3,Zhou,Li_Deng}
and are of great interest due to their application in control
theory, signal processing, complex networks, etc
\cite{Barbosa1,Barbosa2,Pires,Li_Sun,Wang_He,Tang et al}.

In this paper we investigate the possibility to control 
unstable equilibria and unstable periodic orbits 
in fractional order chaotic systems by a time-delayed feedback method. 
Pyragas introduced the time-delayed feedback 
control (TDFC) in 1992 by constructing a control force in a form of
a continuous feedback proportional to the difference between 
the present and an earlier value of an appropriate system 
variable $\eta$, i.e. $K[\eta(t-T)-\eta(t)]$, with $K$ and $T$
being the constant control parameters denoting 
the feedback gain and the time delay, respectively
\cite{SCH08,PYR92,PYR06}. To stabilize unstable
periodic orbits, the delay $T$ is chosen to match 
the period of the unstable orbit. In the case 
of controlling unstable equilibria, the optimal delay is
related to the intrinsic characteristic time scale 
given by the eigenfrequencies of the uncontrolled system.
In both cases, the control force vanishes when the target 
state is reached, rendering the method noninvasive.  
In this sense, the unstable states of the uncontrolled
system are not changed, since the control force acts only 
if the system deviates from the state to be stabilized.

The main advantage of the Pyragas method over the other control methods
is that it does not require the knowledge of the system's equations 
and the positions of the unstable states whose control is desireable.
The method has been used in many concrete applications 
\cite{PT93,BDG94,PBA96,HCT97,SBG97,KF00,FSK02,RP04,LBJ04,PHT05},
and also rigorously investigated analytically 
\cite{HS05,YWH06,DHS07,SSG94,PYR95,AP04,AP05,SS97,P01,PPK02,PPK04,JBO97,NAK97,NU98,FFG07,JFG07,PS07}. 
In recent works \cite{GU08,GU10}, it has been shown that 
the efficiency of the method is greatly improved by deterministic or stochastic 
modulation of the time delay $T$. This variable delay 
feedback control (VDFC) has been shown successful in 
stabilization of unstable equilibria in systems described 
by both ordinary and delay differential equations.

This paper is organized as follows. In Sec. II we present
the time-delayed feedback controller for stabilization of 
unstable equilibria in fractional-order 
dynamical systems. 
We first look at a general non-diagonal case of such a control
scheme and derive the stability conditions for the 
equilibrium points in the presence and absence of control.
To illustrate the control method, we choose a fractional 
order R\"{o}ssler system in a chatoic regime controlled via
a single state variable with a Pyragas-type feedback force.
By performing a linear stability analysis of the controlled 
system, we calculate the domain of successful 
control of the unstable equilibria in the plane parametrized by
the feedback gain and the delay time. It is shown, 
both numerically and analytically, that the unstable 
equilibrium point with an odd number of positive real eigenvalues 
cannot be controlled by the time-delayed feedback method
for any values of the control parameters, thus 
extending the validity of the odd-number limitation theorem
\cite{NAK97,NU98} 
to the case of fractional-order systems. 
In Section III we perform a numerical calculation of the control
domain by the Pyragas delayed feedback with a modulated time 
delay in a form of a sine-wave. 
In Section IV we give a numerical evidence that a 
time-delayed feedback control can successfully stabilize unstable 
periodic orbits, and estimate the corresponding 
feedback gain intervals for which such a control is possible.
A summary of the obtained results is given in Section V.

\section{Delayed feedback control of unstable equilibria}

\subsection{Stability analysis}

We consider a general $n$-dimensional nonlinear fractional-order dynamical 
system under a non-diagonal form of delayed feedback control
in the sense of Pyragas:
\begin{equation}
\left\lbrace
  \begin{array}{l l}
    D_{\ast}^{\alpha_1} x_1(t)&=f_1(\textbf{x}(t))+F_1(t),\\
    D_{\ast}^{\alpha_2} x_2(t)&=f_2(\textbf{x}(t))+F_2(t),\\
	&\vdots\\
    D_{\ast}^{\alpha_n} x_n(t)&=f_n(\textbf{x}(t))+F_n(t),
  \end{array}
\right.
\label{2.1}
\end{equation}
where
\begin{equation}
F_i(t)=\sum_{j=1}^n K_{ij}\left[x_j(t-T)-x_j(t)\right]
\label{2.2}
\end{equation}
is the delayed feedback force applied to the $i$th component
of the system, consisting of contributions of all the system components,
$K_{ij}$ are the gain factors of the feedback terms, 
$T$ is the constant time delay, $\textbf{x}=(x_1,x_2,\dots,x_n)$ is the 
state vector, and $\mathbf{f}=(f_1,f_2,\dots,f_n)$ is the nonlinear 
vector field that determines the dynamics of the unperturbed system. 
The notation $D_*^\alpha$ is the time fractional derivative in the
Caputo sense defined as \cite{Caputo}:
\begin{equation}
D_{\ast}^{\alpha}f(t)= \left\lbrace
  \begin{array}{c l}
    \displaystyle\frac{1}{\Gamma(m-\alpha)}\int_{0}^{t}\frac{f^{(m)}(\tau)}{(t-\tau)^{\alpha+1-m}}\mathrm{d}\tau & 
\text{, $m-1<\alpha<m$},\\
\\
    \displaystyle\frac{\mathrm{d}^mf(t)}{\mathrm{d}t^m} & \text{, $\alpha=m$, $m\in
\mathbb{N^+}$},
  \end{array}
\right.
\label{2.3}
\end{equation}
which is related to the famous Riemann-Liouville
fractional integral \cite{Miller_Ross}
\begin{equation}
J^{\alpha}f(t)=\frac{1}{\Gamma(\alpha)}\int_0^{t}(t-\tau)^{\alpha-1}f(\tau)\mathrm{d}\tau.
\label{2.4}
\end{equation}
i.e.
\begin{equation}
D_{\ast}^{\alpha}f(t)=J^{m-\alpha}\frac{\mathrm{d}^m}{\mathrm{d}t^m}f(t),
\label{2.5}
\end{equation}
where $m=[\alpha]$, i. e., $m$ is the first integer which is not less than
$\alpha$.
In the following, we consider the fractional orders $\alpha_i$ to be
in the interval $(0,1)$.

In the case $K_{ij}=K\delta_{ij}$, where $\delta_{ij}$ is the Kronecker
delta, the generalized control scheme (\ref{2.1}) reduces to TDFC with a 
diagonal coupling, and when the control force is applied only to a
single system component and consists only of contributions of the 
same component, it yields the original TDFC control scheme
introduced by Pyragas.

Let $P=(x_1^*,x_2^*,\dots,x_n^*)$ be an arbitrary 
equilibrium point of the system (\ref{2.1}) in the absence of 
control ($K_{ij}=0$), being a solution to the nonlinear algebraic 
system:
\begin{equation}
\left\lbrace
  \begin{array}{l l}
    f_1(x_1,x_2,\dots,x_n)&=0,\\
    f_2(x_1,x_2,\dots,x_n)&=0,\\
	&\vdots\\
    f_n(x_1,x_2,\dots,x_n)&=0.\\
  \end{array}
\right.
\label{2.6}
\end{equation}
Assuming that $P$ is an unstable equilibrium point of the 
uncontrolled system, we wish to find the domain in the parameter 
space of the feedback gains $K_{ij}$ and the time-delay $T$ 
for which $P$ becomes locally asymptotically stable under TDFC
force (\ref{2.2}).
The stability of $P$ under a non-diagonal feedback control
(\ref{2.1})--(\ref{2.2}) can be determined by linearizing 
(\ref{2.1}) around $P$, which leads to the linear 
autonomous system:
\begin{equation}
\left(
  \begin{array}{c l c}
    D_{\ast}^{\alpha_1} \widetilde{x}_1(t)\\
    D_{\ast}^{\alpha_2} \widetilde{x}_2(t)\\
	\vdots\\
    D_{\ast}^{\alpha_n} \widetilde{x}_n(t)
  \end{array}
\right) = \widehat{\textbf{A}}\cdot\left(
  \begin{array}{c l c}
    \widetilde{x}_1(t)\\
    \widetilde{x}_2(t)\\
	\vdots\\
    \widetilde{x}_n(t)
  \end{array}
\right) +
\left(
  \begin{array}{c}
    \widetilde{F}_1(t)\\
    \widetilde{F}_2(t)\\
	\vdots\\
    \widetilde{F}_n(t)
  \end{array}
\right),
\label{2.7}
\end{equation}
where 
\begin{equation}
\widehat{\textbf{A}}=\left(
  \begin{array}{c c c c}
    a_{11} & a_{12} & \dots & a_{1n}\\
    a_{21} & a_{22} & \dots & a_{2n}\\
    \vdots & \vdots & \ddots & \vdots\\
    a_{n1} & a_{n2} & \dots & a_{nn}\\
  \end{array}
\right)
\label{2.8}
\end{equation}
is the Jacobian matrix of the free-running system, 
with $a_{ij}=(\partial f_i/\partial x_j)$
calculated at $P$, $\widetilde{x}_i(t)=x_i(t)-x_i^\ast$
are the transformed coordinates in which the equilibrium point 
is at the origin, and $\widetilde{F}_i(t)$ are the components of the 
feedback control force in the new coordinates, i.e.
\begin{equation}
\widetilde{F}_i(t)=\sum_{j=1}^n K_{ij}\left[\widetilde{x}_j(t-T)-\widetilde{x}_j(t)\right].
\label{2.9}
\end{equation}
By applying the Laplace transform to Eqs. (\ref{2.7}) and by 
using the formula for the Laplace transform of the
fractional derivative in the Caputo sense \cite{Podlubny}:
\begin{equation}
\mathcal{L}[D_{\ast}^{\alpha}\widetilde{x}_i(t)]=s^\alpha
X_i(s)-\sum_{k=0}^{m-1}\widetilde{x}_i^{(k)}(0+)s^{\alpha-1-k},
\label{2.10}
\end{equation}
where $X_i(s)=\mathcal{L}[\widetilde{x}_i(t)]$ are the Laplace images
and $\widetilde{x}_i^{(k)}(0)$ are the initial conditions, we obtain:
\begin{equation}
\Delta(s)\cdot\mathbf{X}(s)=\mathbf{B}(s),
\label{2.11}
\end{equation}
where 
\begin{widetext}
\begin{equation}
\Delta(s)=\left(
  \begin{array}{c c c c}
    s^{\alpha_1}-a_{11}+K_{11}(1-e^{-sT}) & -a_{12}+K_{12}(1-e^{-sT}) & \dots & -a_{1n}+K_{1n}(1-e^{-sT})\\
    -a_{21}+K_{21}(1-e^{-sT}) & s^{\alpha_2}-a_{22}+K_{22}(1-e^{-sT}) & \dots & -a_{2n}+K_{2n}(1-e^{-sT})\\
    \vdots & \vdots & \ddots & \vdots\\
    -a_{n1}+K_{n1}(1-e^{-sT}) & -a_{n2}+K_{n2}(1-e^{-sT}) & \dots & s^{\alpha_n}-a_{nn}+K_{nn}(1-e^{-sT})
  \end{array}
\right)
\label{2.12}
\end{equation}
\end{widetext}
represents a characteristic matrix of system (\ref{2.7}),
$\mathbf{X}(s)=\textrm{col}(X_1(s),X_2(s),\dots,X_n(s))$ is the
column vector of the Laplace images, and
\begin{eqnarray}
\mathbf{B}(s)
=\left(
  \begin{array}{c}
    \widetilde{x}_1(0)s^{\alpha_1-1}+e^{-sT}\sum_{j=1}^n K_{1j}\int_{-T}^0 \widetilde{x}_j(t)e^{-st}dt\\
    \widetilde{x}_2(0)s^{\alpha_2-1}+e^{-sT}\sum_{j=1}^n K_{2j}\int_{-T}^0 \widetilde{x}_j(t)e^{-st}dt\\
    \vdots \\
    \widetilde{x}_n(0)s^{\alpha_n-1}+e^{-sT}\sum_{j=1}^n K_{nj}\int_{-T}^0 \widetilde{x}_j(t)e^{-st}dt
  \end{array}
\right).\nonumber\\
\label{2.13}
\end{eqnarray}
For the sake of the argument, let us assume that all the roots $s$ of $\textrm{det}[\Delta(s)]=0$
are positioned in the left complex s-plane ($\textrm{Re}(s)<0$).
The assumption is equivalent to the claim that $\textrm{det}[\Delta(s)]\neq0$
for $\textrm{Re}(s)\geq0$, meaning that $\Delta(s)$ is invertible matrix
in the right complex s-plane, and thus 
$\mathbf{X}(s)=\Delta(s)^{-1}\cdot\mathbf{B}(s)$ has a
unique solution $\mathbf{X}(s)$ for $\textrm{Re}(s)\geq0$.
Furthermore, it is easy to check that 
\begin{equation}
\lim_{s\rightarrow 0}s\mathbf{X}(s)=\lim_{s\rightarrow 0}\left[\Delta(s)^{-1}\cdot s\mathbf{B}(s)\right]=\mathbf{0},
\label{2.14}
\end{equation}
from which we have 
\begin{equation}
\lim_{t\rightarrow +\infty}\widetilde{\mathbf{x}}(t)=\lim_{s\rightarrow 0}s\mathbf{X}(s)=\mathbf{0}.
\label{2.15}
\end{equation}
by using the final-value theorem for the Laplace transform. 
Thus, we prove the following theorem.

\textbf{Theorem 1.} The equilibrium point $P$ of the 
system (\ref{2.1})--(\ref{2.2}) is locally asymptotically
stable if and only if all the roots $s$ of the 
characteristic equation:
\begin{equation}
\det\left[\Delta(s)\right]=0,
\label{2.16}
\end{equation}
have negative real parts, i.e. 
\begin{equation}
|\arg(s)|>\pi/2.
\label{2.17}
\end{equation}
The matrix $\Delta(s)$ is given by Eq. (\ref{2.12}),
and its components $a_{ij}=(\partial f_i/\partial x_j)$
are evaluated at the equilibrium point $P$. \hfill $\Box$

\textit{Remark 1.} For the case when there is no control ($K_{ij}=0$),
the conditions for stability of a particular equilibrium point 
is still provided by Theorem 1. In this case, 
the characteristic matrix of the uncontrolled system
simplifies to 
\begin{equation}
\Delta(s)=\mathbf{S}\cdot\mathbf{\widehat{I}}-\mathbf{\widehat{A}},
\label{2.18}
\end{equation}
where $\mathbf{S}=\textrm{col}(s^{\alpha_1},s^{\alpha_2},\dots,s^{\alpha_n})$,
$\mathbf{\widehat{I}}$ is $n\times n$ identity matrix, and $\mathbf{\widehat{A}}$
is the Jacobian matrix of the unperturbed system evaluated at
the equilibrium point, and given by Eq. (\ref{2.8}).
In the special case $\alpha_1=\alpha_2=\dots=\alpha_n=\alpha$,
the characteristic Eq. (\ref{2.18}) can be recast into the form
\begin{equation}
\det\left(\lambda\mathbf{\widehat{I}}-\mathbf{\widehat{A}}\right)=0,
\label{2.19}
\end{equation}
which is a polynomial of degree $n$ in $\lambda$, where $\lambda=s^\alpha$. 
In this case, the stability condition (\ref{2.17}) can be rewritten 
in the form of Matignon \cite{Matignon}
\begin{equation}
|\arg(\lambda)|>\alpha\pi/2,
\label{2.20}
\end{equation}
meaning that the stability region is bounded by a cone, with vertex
at the origin, extending into the right half of the complex $\lambda$-plane
such that it encloses an angle of $\pm \alpha\pi/2$ with the positive
real axis. Therefore, the equilibrium point of the unperturbed system
is stable if and only if all the roots of the characteristic polynomial 
(\ref{2.19}) are placed outside this cone.

To derive another important result related to the limitation
of the time-delayed feedback method (\ref{2.1})--(\ref{2.2}), 
we consider the function $\det[\Delta(s)]$, with $\Delta(s)$
given by Eq. (\ref{2.12}), and $s\in\mathbb R^+$. 
We take $P$ to be an unstable equilibrium of (\ref{2.1})
in the absence of external perturbation ($K_{ij}=0$),
and $\mathbf{\widehat{A}}$ the Jacobian matrix of $P$.
One can easily deduce that
\begin{equation}
\lim_{s\rightarrow+\infty}\det[\Delta(s)]=+\infty,
\label{2.21}
\end{equation}
and
\begin{equation}
\lim_{s\rightarrow0^+}\det[\Delta(s)]=\det[-\mathbf{\widehat{A}}]=\prod_{i=1}^n(-e_i),
\label{2.22}
\end{equation}
where $e_i$ are the eigenvalues of $\mathbf{\widehat{A}}$.
Evidently, if $\widehat{\mathbf{A}}$ has an odd number of positive 
real eigenvalues, then $\lim_{s\rightarrow0^+}\det[\Delta(s)]<0$. 
In this case, the sign of $\det[\Delta(s)]$ is changed
from negative to positive when $s$ sweeps the real interval $[0,+\infty)$.
Since $\det[\Delta(s)]$ is a smooth function in $s$, there 
exists at least one positive real root of the 
characteristic equation $\det[\Delta(s)]=0$, meaning that the 
equilibrium point $P$ cannot be stabilized by the time-delayed
feedback controller (\ref{2.2}). The result is summarized 
in the following theorem.

\textbf{Theorem 2. (odd-number limitation)}
Let $P$ be an unstable equilibrium point of the fractional-order 
system (\ref{2.1}) in the absence of control ($K_{ij}=0$),
and $\mathbf{\widehat{A}}$ the corresponding Jacobian matrix
at $P$. If $\mathbf{\widehat{A}}$ has an odd number of 
positive real eigenvalues, then the time-delayed feedback 
control (\ref{2.2}) cannot stabilize the unstable 
equilibrium $P$ for any values of the control parameters 
$K_{ij}$ and $T$. \hfill $\Box$

The result is an extension of the odd-number limitation 
theorem \cite{NAK97,NU98} to fractional-order systems with respect to unstable
fixed points. We note that the odd-number limitation  
has recently been refuted by Fiedler et al. \cite{FFG07,JFG07} 
for the case of unstable periodic orbits in systems described by 
ordinary differental equations.

\subsection{Numerical example}

To illustrate the time-delayed feedback control in fractional 
order chaotic systems, we consider a fractional order R\"{o}ssler 
system in the form:
\begin{equation}
\left\lbrace
  \begin{array}{c l c}
    D_{\ast}^\alpha x(t)&=-y(t)-z(t),\\
    D_{\ast}^\alpha y(t)&=x(t)+ay(t)+F(t),\\
    D_{\ast}^\alpha z(t)&=z(t)\left[x(t)-c\right]+b,
  \end{array}
\right.
\label{2.23}
\end{equation}
where
\begin{equation}
F(t)=K[y(t-T)-y(t)]
\label{2.24}
\end{equation}
is the Pyragas feedback controller applied through a
single component ($y$-channel), and $a$, $b$ and $c$ are 
the parameters of the free-running system.
In the following, we take $a=0.4$, $b=0.2$, $c=10$ 
and $\alpha=0.9$, for which the uncontrolled system 
$(K=0)$ has a chaotic attractor (see Fig. 1) \cite{Li_Chen}.

\begin{figure*}
\includegraphics[width=0.8\textwidth,height=!]{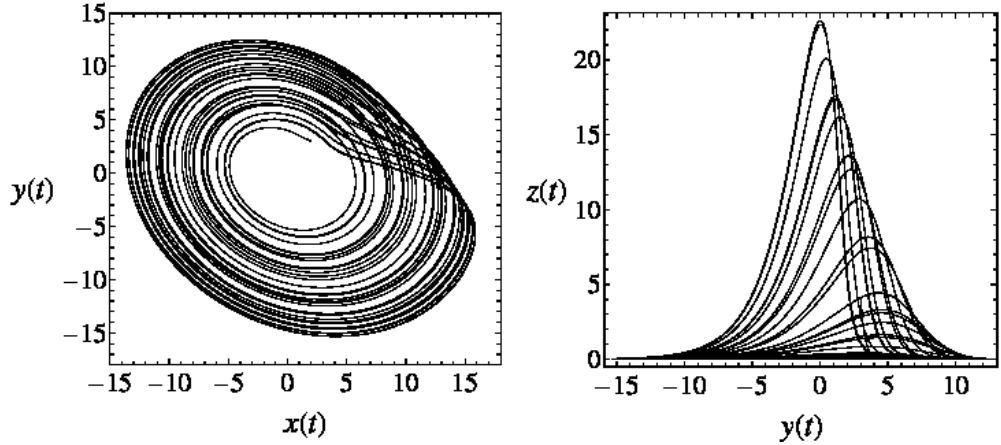}
\caption{The phase plots of the chaotic attractor in the 
free-running fractional-order R\"ossler system. The parameters 
are $a=0.4$, $b=0.2$, $c=10$ and $\alpha=0.9$.}
\end{figure*}

The unperturbed R\"{o}ssler system has two equilibrium points 
$P_1=(x_+^\ast,y_+^\ast,z_+^\ast)$ and $P_2=(x_-^\ast,y_-^\ast,z_-^\ast)$, where
\begin{eqnarray}
x_{\pm}^\ast&=&\frac{c}{2}\left(1\pm\sqrt{1-\frac{4ab}{c^2}}\right),\\
y_{\pm}^\ast&=&-\frac{c}{2a}\left(1\pm\sqrt{1-\frac{4ab}{c^2}}\right),\\
z_{\pm}^\ast&=&\frac{c}{2a}\left(1\pm\sqrt{1-\frac{4ab}{c^2}}\right).
\label{2.25}
\end{eqnarray}
Linearization around the equilibrium points leads to the following 
linear autonomous system:
\begin{equation}
\left(
  \begin{array}{c l c}
    D_{\ast}^\alpha \widetilde{x}(t)\\
    D_{\ast}^\alpha \widetilde{y}(t)\\
    D_{\ast}^\alpha \widetilde{z}(t)
  \end{array}
\right) = \widehat{\textbf{A}}\cdot\left(
  \begin{array}{c l c}
    \widetilde{x}(t)\\
    \widetilde{y}(t)\\
    \widetilde{z}(t)
  \end{array}
\right),
\label{2.26}
\end{equation}
where 
\begin{equation}
\widehat{\textbf{A}}=\left(
  \begin{array}{c l c}
    0 & -1 & -1\\
    1 & a & 0\\
    z_\pm^\ast & 0 & x_\pm^\ast-c
  \end{array}
\right)
\label{2.27}
\end{equation}
is the Jacobian matrix, and $\widetilde{x}(t)=x(t)-x_\pm^\ast$,
$\widetilde{y}(t)=y(t)-y_\pm^\ast$, $\widetilde{z}(t)=z(t)-z_\pm^\ast$
are the transformed coordinates in which the corresponding fixed point 
is at the origin. According to Eq. (\ref{2.20}), the equilibrium 
point of the linearized system (\ref{2.26})
is asymptotically stable if and only if $|\arg(\lambda)|>\alpha\pi/2$ for
all the eigenvalues $\lambda$ of the Jacobian matrix $\widehat{\textbf{A}}$. 
The equilibrium point 
$P_1=(x_+^\ast,y_+^\ast,z_+^\ast)=(9.9919,-24.98,24.98)$ has eigenvalues
$\lambda_1=0.3844$ and $\lambda_{2,3}=0.0038\pm5.0964i$. It is an unstable saddle
point of index 1 since $\lambda_1>0$,
$|\arg(\lambda_{2,3})|=1.5701>\alpha\pi/2$. The equilibrium point
$P_2=(x_-^\ast,y_-^\ast,z_-^\ast)=(0.008,-0.02,0.02)$ 
has eigenvalues $\lambda_1=-9.9900$ and $\lambda_{2,3}=0.1990\pm0.9797i$,
and it is an unstable saddle point of index 2 since $\lambda_1<0$,
$|\arg(\lambda_{2,3})|=1.3704<\alpha\pi/2$ \cite{Deng_Lu}.

In the presence of TDFC, the linearized version of the system (\ref{2.23}) 
around the equilibrium points states:
\begin{equation}
\left(
  \begin{array}{c l c}
    D_{\ast}^\alpha \widetilde{x}(t)\\
    D_{\ast}^\alpha \widetilde{y}(t)\\
    D_{\ast}^\alpha \widetilde{z}(t)
  \end{array}
\right) = \widehat{\textbf{A}}\cdot\left(
  \begin{array}{c l c}
    \widetilde{x}(t)\\
    \widetilde{y}(t)\\
    \widetilde{z}(t)
  \end{array}
\right)+K\left(
  \begin{array}{c l c}
    0\\
    \widetilde{y}(t-T)-\widetilde{y}(t)\\
    0
  \end{array}
\right).\label{2.28}
\end{equation}
According to Theorem 1, the zero solution of system (\ref{2.28}) 
is asymptotically stable if and only if all the
roots $s$ of the characteristic equation:
\begin{equation}
\det\left[\Delta(s)\right]=0
\label{2.29}
\end{equation}
have negative real parts, i.e. $|\arg(s)|>\pi/2$, where
the characteristic matrix $\Delta(s)$ is given by:
\begin{equation}
\Delta(s)=\left(
\begin{array}{ccc}
\ s^\alpha & 1 & 1\\
{-1} & s^\alpha-a+K\left(1-e^{-sT}\right) & 0\\
{-z}_\pm^\ast & 0 & s^\alpha-(x_\pm^\ast-c)
\end{array}
\right).
\label{2.30}
\end{equation}
The characteristic Eq. (\ref{2.29}) can be numerically analyzed 
to obtain the domains of control for the unstable steady states 
$P_{1,2}$ in the plane parametrized by the feedback gain $K$ and 
the time delay $T$. 

In the absence of control, the equilibrium point $P_1$ has an
odd number (one) of positive real eigenvalues, and according
to Theorem 2, it cannot be stabilized by the TDFC method. 
This result has been confirmed by a numerical analysis of 
the characteristic Eq. (\ref{2.29}), showing absence of 
stability domain in the $(K,T)$ parameter plane. 
This observation is further confirmed by a numerical simulation 
of the system (\ref{2.23}) under TDFC (\ref{2.24}).
On the other hand, the fixed point $P_2$ can be controlled by TDFC,
and the resulting stability domain is shown in Fig. 2.
\begin{figure}
\includegraphics[width=0.6\columnwidth,height=!]{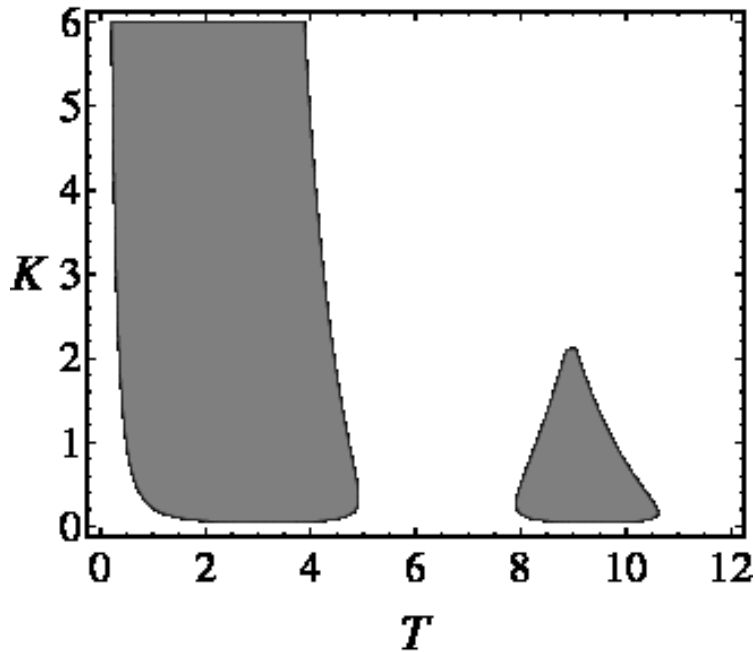}
\caption{The stability domain in the $(K,T)$ parametric plane 
of the unstable equilibrium $P_2$ in the fractional-order 
R\"ossler system (\ref{2.23}) under time-delayed feedback control
(\ref{2.24}). 
The shaded areas correspond to the control parameter 
values for which stabilization of the fixed point $P_2$ is 
achievable. The parameters of the free-running system
are $a=0.4$, $b=0.2$, $c=10$ and $\alpha=0.9$.}
\end{figure}
The stability islands (shaded areas) denote the values of
the control parameters $K$ and $T$ for which all the eigenvalues
$s$ of the characteristic Eq. (\ref{2.29}) are lying on the
left complex $s$-plane, thus satisfying the stability condition 
(\ref{2.17}).
For these values of the control parameters, the control of the 
fixed point $P_2$ is successful. As a verification, 
we performed a computer simulation of TDFC by numerically
integrating the system (\ref{2.23})--(\ref{2.24}). 
The resulting diagrams are shown in Fig. 3. The simulations
were done by using a predictor-corrector 
Adams-Bashford-Moulton numerical scheme for solving fractional order 
differential equations \cite{Diethelm_Ford}. Panels (a),
(b) and (c) depict the dynamics of the state variables
$x(t)$, $y(t)$ and $z(t)$, respectively, and panel (d) 
shows the corresponding time series of the control signal 
$F(t)$. In the simulations,
the control parameters were $K=2$ and $T=3$, belonging
to the domain of successful TDFC control depicted in Fig. 2. 
As expected, the simulation confirms a successful stabilization 
of the unstable equilibrium $P_2$. Moreover, as indicated from 
panel (d) in Fig. 3, the control signal $F(t)$ vanishes when 
the control is achieved, meaning that the control scheme 
is noninvasive.

We note that the above analysis has been repeated for 
different parameter values of the free-running system. 
In each case, the resulting stability domains
computed from Eqs. (\ref{2.29})--(\ref{2.30})
are in agreement with the numerical simulation of 
the TDFC method.
Specifically, for $a=0.4$, $b=0.2$, $c=10$ and 
variable $\alpha$, we observed a decrease in the 
stability region as $\alpha$
is increased from $\alpha=0.9$ to $\alpha=1$. 
On the other hand, as $\alpha$ becomes
smaller than $0.9$, the complex-conjugate 
eigenvalues of the equilibrium point $P_2$ eventually 
escape the instability region described by the
Matignon formula (\ref{2.20}), resulting in 
a stable equilibrium $P_2$ even without control.
The critical value $\alpha=\alpha_c$ that corresponds to this
eigenvalue-crossing of the conic surface between the different stability 
regions can be calculated from
Eq. (\ref{2.20}). In this case, $\alpha_c=0.8724$.

\begin{figure*}
\includegraphics[width=0.8\textwidth,height=!]{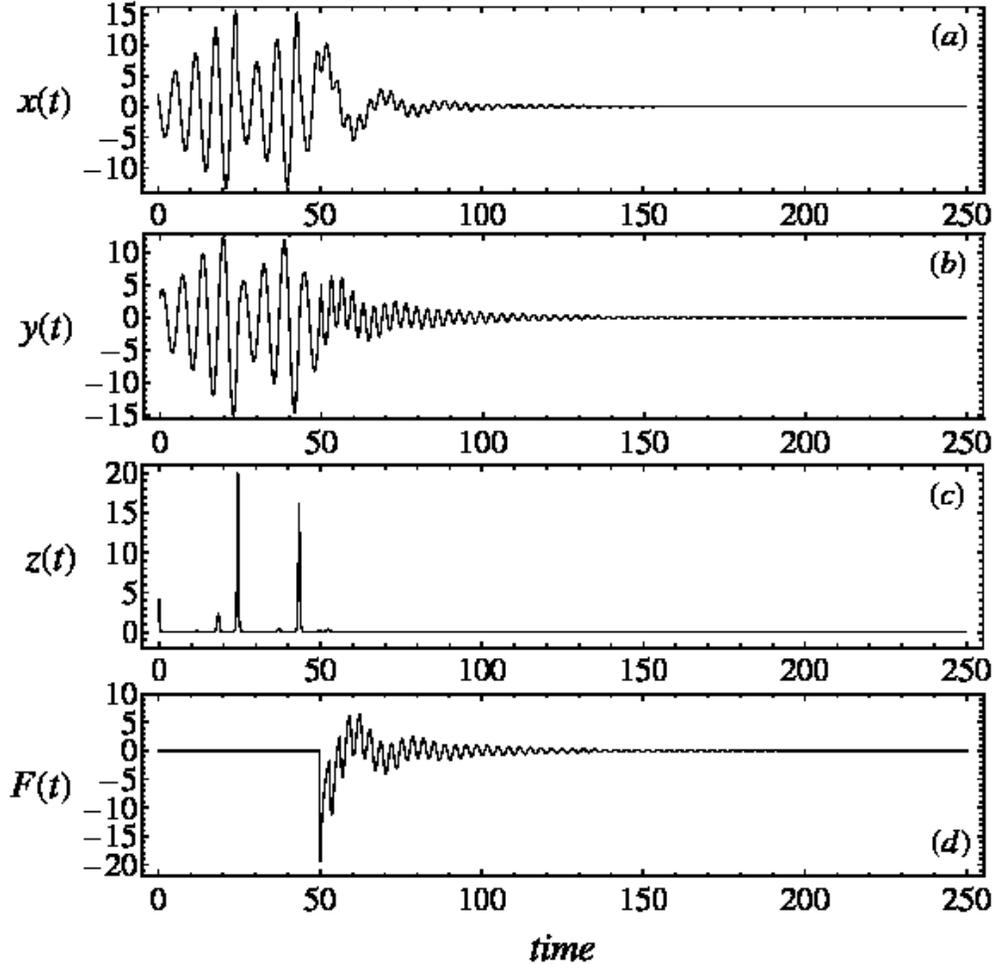}
\caption{Successful stabilization of the unstable equilibrium 
$P_2$ in the fractional-order R\"ossler system by a time-delayed 
feedback control. (a)--(c) Time plots of the state variables 
$x(t)$, $y(t)$ and $z(t)$. (d) The feedback control force $F(t)$
vanishes when the control is achieved.
The uncontrolled fractional-order R\"ossler system 
is in a chaotic regime, with parameters: $a=0.4$, $b=0.2$, 
$c=10$ and $\alpha=0.9$.
The control parameters were: $K=2$, $T=3$. The control
was activated at $t=50$. The total time span shown in each
panel is 250 time units.}
\end{figure*}

\section{Variable-delay feedback control of unstable equilibria}

In recent papers \cite{GU08,GU10}, it has been demonstrated, 
both numerically and 
analytically, that the original Pyragas TDFC scheme can be 
improved significantly by modulating the time delay in an 
$\varepsilon$ interval around some nominal delay value $T_0$.  
In both deterministic and stochastic variants of such a delay 
variation, the stability domain was considerably changed, 
resulting in an extension of the stability area in the 
control parameter space if appropriate modulation is chosen.
In the following, we will demonstrate numerically 
the successfulness of this variable-delay feedback control 
in the case of fractional-order chaotic R\"ossler system (\ref{2.23})
with $a=0.4$, $b=0.2$, $c=10$ and $\alpha=0.9$.
In this case, the feedback force $F(t)$ is given by:
\begin{equation}
F(t)=K[y(t-T(t))-y(t)],
\label{3.1}
\end{equation}
where we choose a time-varyng delay $T(t)$ in a form
\begin{equation}
T(t)=T_0+\varepsilon\sin(\omega t),
\label{3.2}
\end{equation}
modulated around a nominal delay value $T_0$ with 
a sine-wave modulation of amplitude $\varepsilon$ and 
frequency $\omega$. Obviously, if $\varepsilon=0$ then
$T(t)=T_0=\mathrm{const}$, and the variable-delay 
feedback control is reduced to the classical Pyragas 
TDFC scheme.
With this choice of the feedback
force $F(t)$, the control parameters of the proposed 
variable-delay scheme are $K$, $T_0$, $\varepsilon$ and $\omega$,
and thus, the control parameter space is four-dimensional.
For visualisation purposes, we may fix two of the control
parameters and investigate the stability domains in 
the parametric plane spanned by the remaining two 
control parameters. To demonstrate the superiority of 
variable-delay feedback control over TDFC, we fix the 
modulation amplitude $\varepsilon$ and the frequency 
$\omega$ and investigate the control domain in 
$(K,T_0)$ parameter space. Numerical simulations show
that the stability area is gradually increasing as
$\varepsilon$ is increased from zero. Figure 4 shows such a control 
domain for $\varepsilon=1$ and $\omega=10$.
The grey region indicates those values of the 
control parameters $K$ and $T_0$ for which the control 
of the unstable equilibrium $P_2$
is successful. The stability domain is obtained by numerically 
integrating the linearized system 
\begin{equation}
\left(
  \begin{array}{c l c}
    D_{\ast}^\alpha \widetilde{x}(t)\\
    D_{\ast}^\alpha \widetilde{y}(t)\\
    D_{\ast}^\alpha \widetilde{z}(t)
  \end{array}
\right) = \widehat{\textbf{A}}\cdot\left(
  \begin{array}{c l c}
    \widetilde{x}(t)\\
    \widetilde{y}(t)\\
    \widetilde{z}(t)
  \end{array}
\right)+K\left(
  \begin{array}{c l c}
    0\\
    \widetilde{y}(t-T(t))-\widetilde{y}(t)\\
    0
  \end{array}
\right),\label{3.3}
\end{equation}
with the Jacobian matrix $\mathbf{\widehat{A}}$ given
by Eq. (\ref{2.27}).
It is evident that the control 
domain is significantly enlarged in comparison to 
the one in TDFC in Fig. 2. We note that numerical
integration of the system (\ref{3.3}) for the 
unstable equilibrium $P_1$ shows failure
of the variable-delay control scheme for any values 
of $K$ and $T_0$, suggesting validity of the odd-number 
limitation theorem also in the case of a time-varying delay.

\begin{figure}
\includegraphics[width=0.6\columnwidth,height=!]{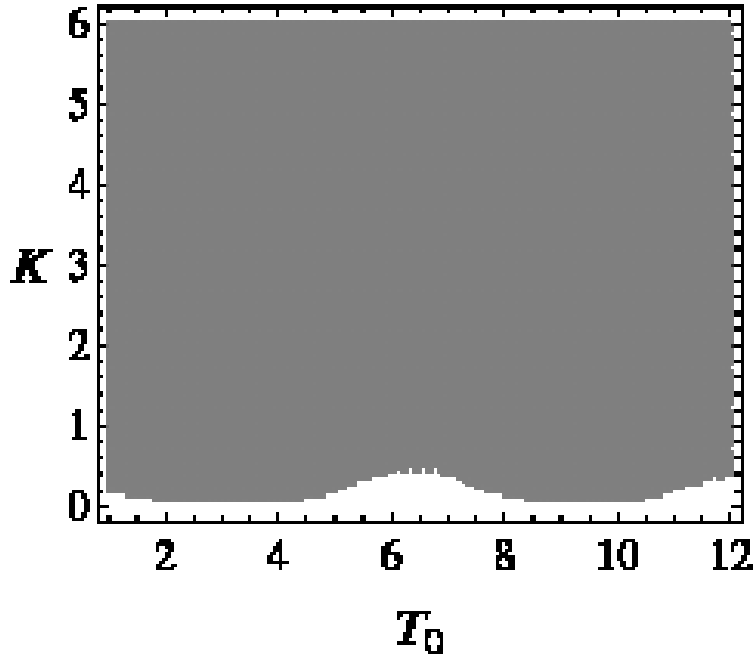}
\caption{Domain of successful variable-delay feedback control in 
the $(K,T_0)$ plane for the unstable equilibrium $P_2$ in the 
chaotic fractional-order R\"ossler system ($a=0.4$, $b=0.2$, 
$c=10$, $\alpha=0.9$). The control delay $T(t)$ is modulated 
with a sine-wave, with $\varepsilon=1$ and $\omega=10$.
Combinations of $K$ and $T_0$ where the control scheme successfully 
stabilizes the fixed point $P_2$ are plotted in grey.
Note the shift of the origin along the $T_0$ axis by an amount 
equal to $\varepsilon$ due to the physical limitation 
$T_0\geq\varepsilon$ of the controller.}
\end{figure}

As a demonstration of the variable-delay feedback control
in the fractional-order R\"ossler system, in panels (a)--(c) of 
Fig. 5 we show the dynamics of the state variables $x(t)$, 
$y(t)$ and $z(t)$ for $K=3$ and $T_0=7$, fixing the modulation 
amplitude $\varepsilon=2$ and frequency $\omega=10$. 
The time series indicate a successful stabilization of the 
unstable equilibrium $P_2$. The method of control is again noninvasive,
as indicated by the vanishing feedback force $F(t)$
in panel (d) of Fig. 5. We note that for these parameter 
values, the control via TDFC $(\varepsilon=0)$ is unsuccessful,
as can be perceived from the stability domain in the TDFC case
depicted in Fig. 2.

\begin{figure*}
\includegraphics[width=0.8\textwidth,height=!]{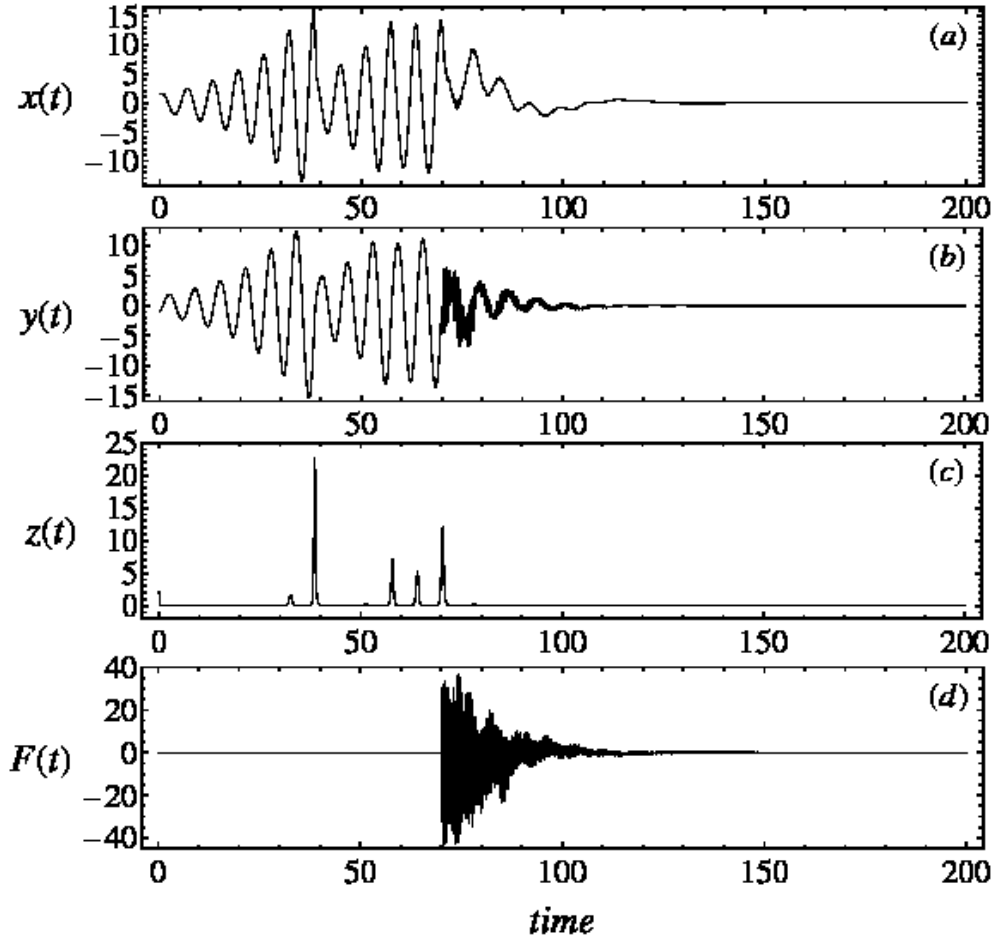}
\caption{A simulation of the variable-delay feedback control
in the chaotic fractional-order R\"ossler system ($a=0.4$, $b=0.2$, 
$c=10$, $\alpha=0.9$). The delay
modulation is in a form of a sine-wave with amplitude $\varepsilon=2$
and frequency $\omega=10$. The time series of the variables
$x(t)$, $y(t)$ and $z(t)$ depicted in panels (a)--(c) indicate a successful
control of the unstable equilibrium $P_2$. The vanishing 
feedback force $F(t)$ in panel (d) shows the noninvasiveness of the
control method. 
The control parameters were: $K=3$, $T_0=7$. The control
was activated at $t=70$. The total time span shown in each
panel is 200 time units.}
\end{figure*}

\section{Delayed feedback control of unstable periodic orbits}

The Pyragas delayed feedback control method was originally
aimed to stabilize unstable periodic orbits embedded into
the chaotic attractor of the free-running system \cite{PYR92}. 
For this purpose, the time delay in the feedback loop was chosen 
to coincide with the period of the target orbit. 
By tuning the feedback gain to an appropriate value,
the stabilization is achieved and the controller perturbation 
vanishes, leaving the target orbit and its period unaltered. 

In this section, we will give a brief demonstration 
of the Pyragas method to 
control unstable periodic orbits in the fractional-order 
R\"ossler system (\ref{2.23})--(\ref{2.24}).
As in the previous discussion, we use $a=0.4$, $b=0.2$, 
$c=10$ and $\alpha=0.9$, for which the system is chaotic
in the absence of external perturbation.

In order to estimate the periods of the unstable orbits
which are typically not known a priori, 
we use the fact that the signal difference
$\mathcal{F}(t)=y(t-\tau)-y(t)$ at a successful 
control asymptotically tends to zero if the 
delay $\tau$ of the controller is
adjusted to match the period $T$ of the target orbit. 
The method consists of calculating the 
dispersion $\left<\mathcal{F}^2\right>$ 
of the control signal at a fixed value
of the feedback gain $K$ for a given range of values 
of the delay $\tau$, excluding the transient period
\cite{PYR92,BJS05}. 
The resulting logarithmic plots of the dependence of the
dispersion $\left<\mathcal{F}^2\right>$ on the delay $\tau$
may contain several segments of finite $\tau$-intervals 
for which $\left<\mathcal{F}^2\right>$ is practically zero,
and a sequence of isolated resonance peaks 
with very deep minima.
The former correspond to the stability domain of 
the fixed point $P_2$, and the latter are the points
at which $\tau$ coincides with some accuracy to the periods 
of the unstable periodic orbits in the 
original system. The estimated values of the 
periods $T$ can be made more accurate if one repeats 
this ``spectroscopy'' procedure for a larger sampling resolution
of the $\tau$ interval encompassing the resonance peaks.
In this way, we have obtained the periods of the 
unstable period-one, period-two, and period-three 
orbits: $T_1\approx6.2$, $T_2\approx12.49$ and $T_3\approx18.89$.
The plot of the dispersion $\left<\mathcal{F}^2\right>$
vs the delay $\tau$ for $K=0.2$ is shown in panel (a) of Fig. 6.

\begin{figure*}
\includegraphics[width=0.8\textwidth,height=!]{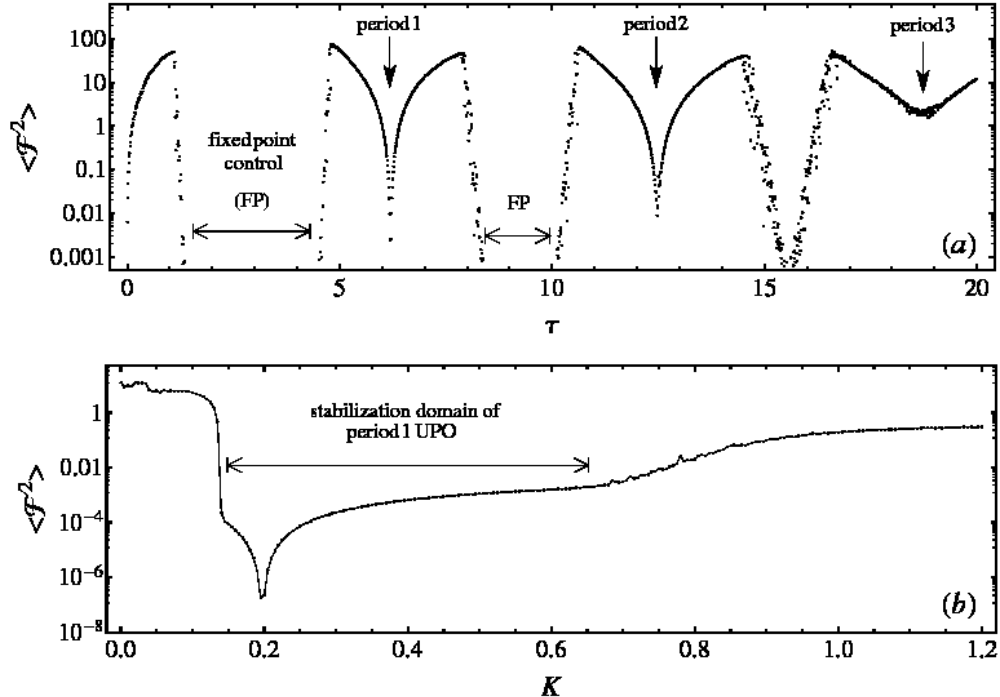}
\caption{Example of the ``spectroscopic'' procedure to 
determine the periods of unstable periodic orbits 
and their control domains under TDFC in the fractional-order 
R\"{o}ssler system ($a=0.4$, $b=0.2$, $c=10$, $\alpha=0.9$). 
(a) The dependence of the dispersion of the control signal 
$\mathcal{F}(t)=y(t-\tau)-y(t)$
upon the time delay $\tau$ for $K=0.2$ reveals segments of finite length
(FP) related to a fixed point stabilization (compare with the $T$-intervals 
in Fig. 2 at $K=0.2$), 
and a resonance peaks at those values of $\tau$ corresponding to the periods of 
unstable periodic orbits. The period-three resonance
point becomes more pronounced for smaller values of $K$.
(b) Calculation of the stability domain for the period-one orbit.
Note the peak at $K\approx0.195$, for which the control of the orbit is the
most robust.}
\end{figure*}

The same approach could be used to calculate
the intervals of the feedback gain $K$ for which the corresponding
orbits can be stabilized with the Pyragas controller.
In panel (b) of Fig. 6 we depict the dependence of the dispersion
$\left<\mathcal{F}^2\right>$ on the feedback gain $K$
when the delay time $\tau$ coincides with the 
period of the first unstable periodic orbit $\tau=T_1=6.2$.
In this case, the interval of the parameter $K$ for which 
the orbit is stabilized is estimated to be 
$K=[0.14,0.65]$. A similar analysis yields 
$K=[0.1,0.22]$ for a period-two, and $K=[0.06,0.15]$
for a period-three orbit. 
It is observed that the control interval of the feedback 
gain $K$ becomes narrower as the period of the target orbit
is increased. 

In Fig. 7 we show the results of the stabilization of period-one 
orbit ($T=6.2$) for $K=0.25$. Panels (a) and (b) show the projection of the
system trajectory in $xy$ and $xz$ planes, respectively,
after the control of the target period-one orbit has been
established. The time-series of the state variables are given 
in panels (c)--(e), and panel (f) shows the feedback
force that vanish after the controller is switched-on,
warranting a noninvasiveness of the control procedure.
Analogous results related to stabilization of period-two
and period-three orbits are given in Figs. 8 and 9.

\begin{figure*}
\includegraphics[width=0.8\textwidth,height=!]{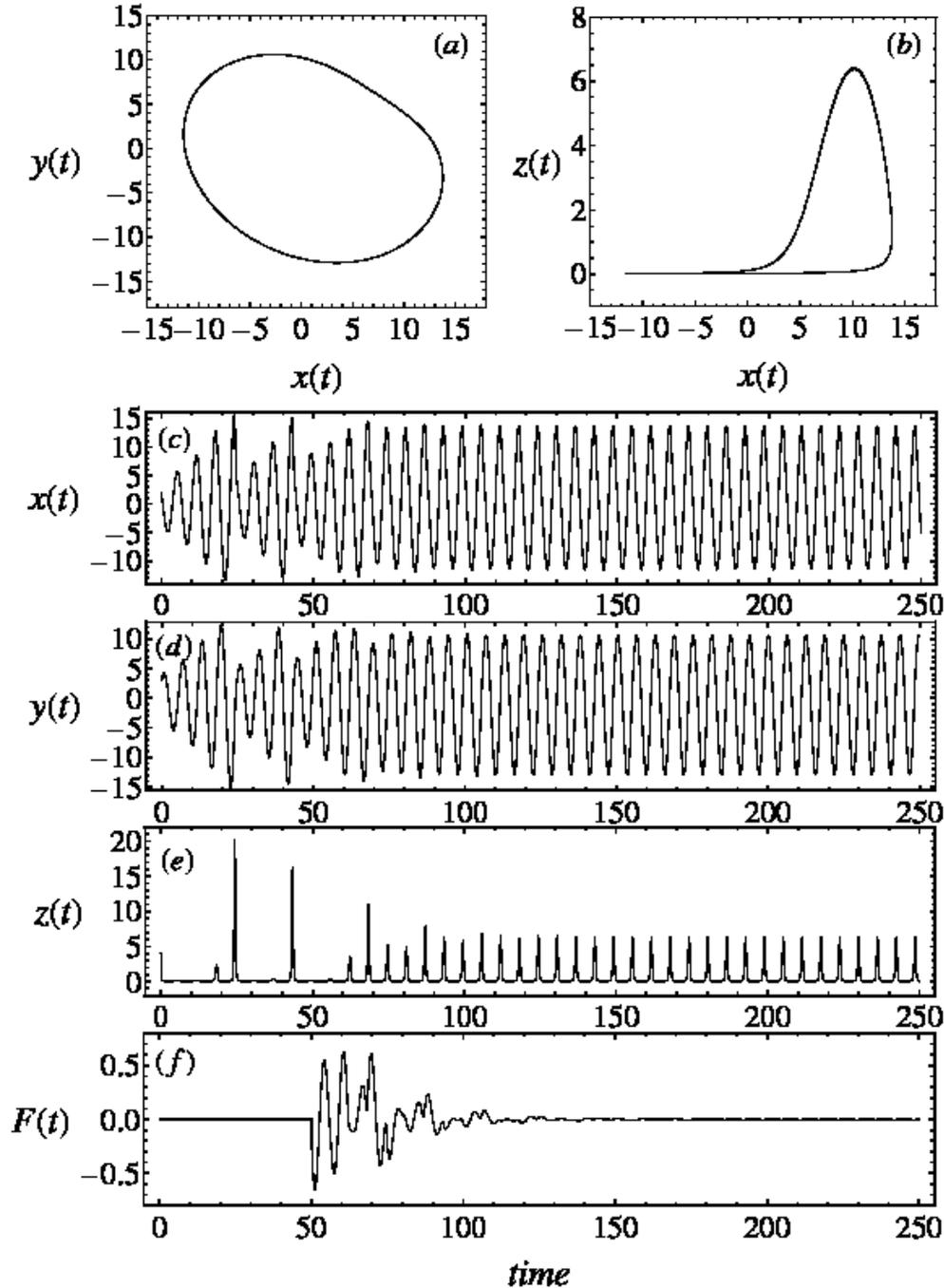}
\caption{A successful stabilization of period-one orbit 
in the chaotic fractional-order R\"{o}ssler system 
by a time-delayed feedback control. The control parameters are
$K=0.25$ and $T=6.2$. 
Panels (a)--(b) depict the phase plots of the controlled periodic orbits 
in the post-transient regime.
Panels (c)--(e) depict the time series of the state
variables $x(t)$, $y(t)$ and $z(t)$. The vanishing 
feedback force $F(t)$ in panel (f) demonstrates 
the noninvasiveness of the control method. 
The parameters of the unperturbed system are $a=0.4$, $b=0.2$, $c=10$ and $\alpha=0.9$.
The control was activated at $t=50$. The total time span shown in each
panel is 250 time units.}
\end{figure*}

\begin{figure*}
\includegraphics[width=0.8\textwidth,height=!]{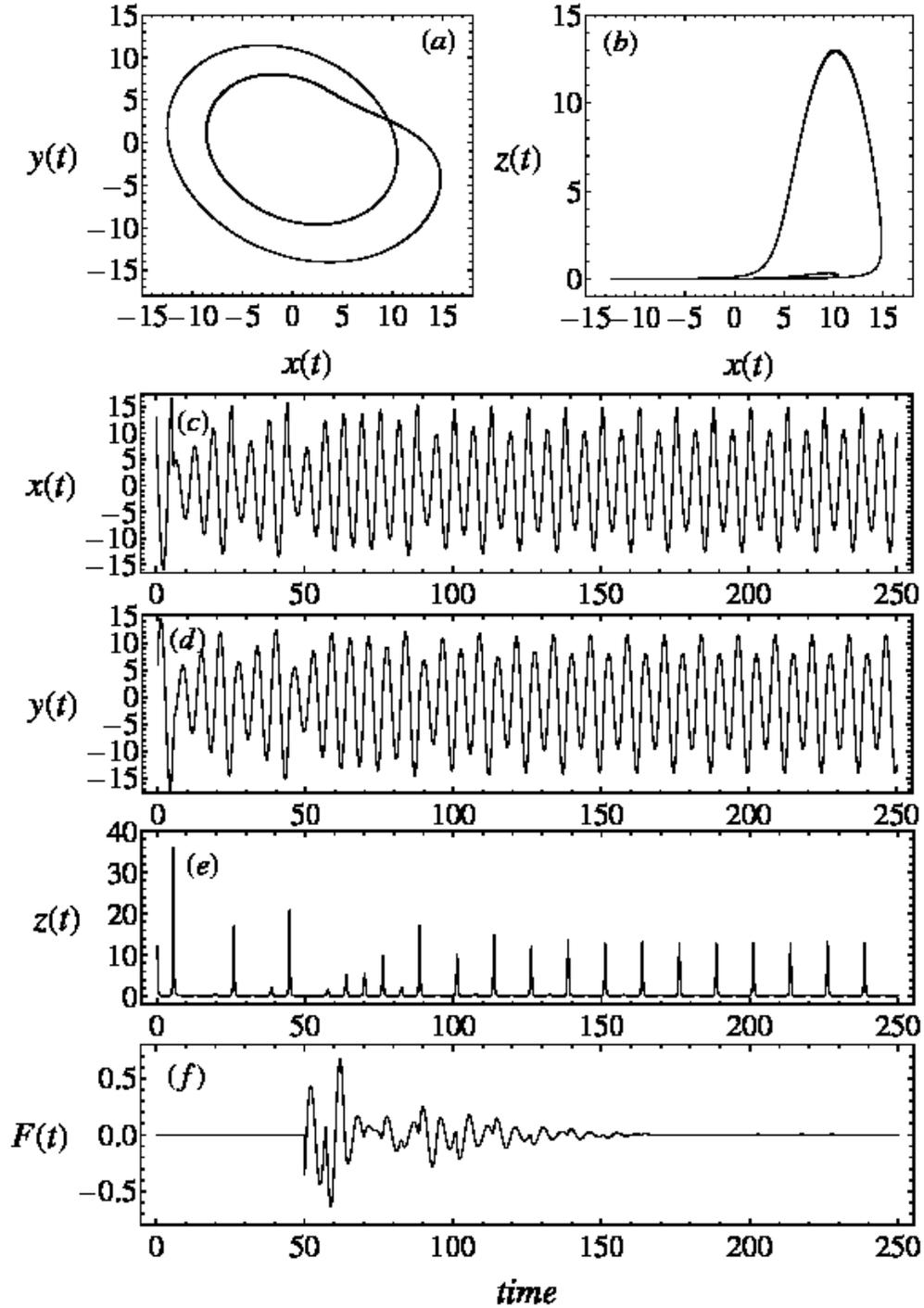}
\caption{Results of  stabilization of period-two orbit 
in the chaotic fractional-order R\"{o}ssler system. 
The control parameters are $K=0.12$ and $T=12.49$. 
The parameters of the unperturbed system are as in Fig. 7. 
The control was activated at $t=50$. The total time span shown in each
panel is 250 time units. Note that one of the period-two peaks
in panel (e) is barely visible.}
\end{figure*}

\begin{figure*}
\includegraphics[width=0.8\textwidth,height=!]{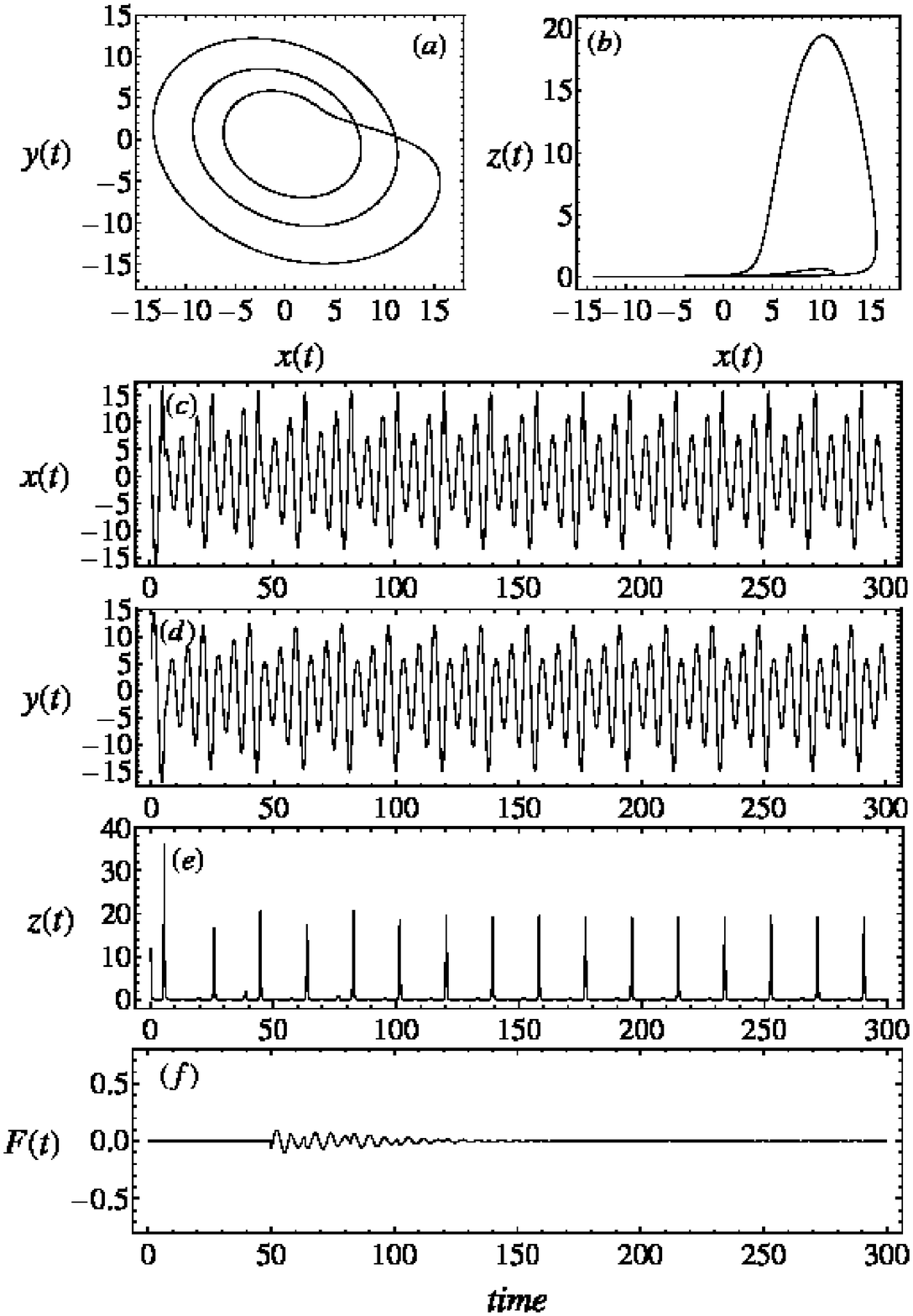}
\caption{Results of  stabilization of period-three orbit 
in the chaotic fractional-order R\"{o}ssler system. 
The control parameters are $K=0.08$ and $T=18.89$. 
The parameters of the unperturbed system are as in Fig. 7. 
The control was activated at $t=50$. The total time span shown in each
panel is 300 time units. Note that one of the period-three peaks
in panel (e) is too small to be seen on the scale of this figure.}
\end{figure*}

A detailed bifurcation analysis of the 
chaotic R\"{o}ssler system described by ordinary 
differential equations and subjected to a time-delayed feedback 
control has been performed recently \cite{BJS05}, revealing 
multistability and a large variety of different attractors
that are not present in the free-running system. A similar analysis
in the case of fractional-order chaotic systems is left for future studies.

\section{Summary and conclusions}
We have shown that the time-delayed feedback control 
can be used to stabilize unstable steady states and 
unstable periodic orbits in fractional-order chaotic systems. 
Although the control method was illustrated specifically 
for the fractional order R\"{o}ssler system, it has also 
successfully been applied to stabilize unstable equilibria 
and unstable periodic orbits in various other fractional-order 
dynamical systems. In all the cases, delayed feedback control with 
a variable time-delay significantly enlarges the stability 
region of the steady states in comparison to 
the classical Pyragas TDFC scheme with a constant delay.

We find that equilibrium points that have an odd number
of positive real eigenvalues cannot be stabilized by TDFC for any 
values of the feedback control parameters. The result is known as 
the odd-number limitation theorem, which extends to the
case of fractional-order systems, as purported by Theorem 2. 
The odd-number limitation is also confirmed numerically in the case of a 
variable-delay feedback control.

An analytical treatment of delayed feedback control of unstable
periodic orbits in fractional-order systems is still lacking, and 
constitutes a promising subject for a future research. 
Applying the extended versions of the delayed feedback controller
\cite{SSG94,AP04} to fractional-order systems is another interesting topic
not tackled in this paper. This is especially important regarding the
observations for the system used in this paper, that the control 
domains are becoming smaller for higher orbits, such that the 
periodic orbits of periods higher than three
practically cannot be stabilized by the original 
controller.

A detailed analysis of the variable-delay feedback control in 
fractional-order systems, including a theoretical understanding 
of the method and numerical computation of the stability domains 
in different parameter planes and for different types of delay 
modulations are also left for future studies.


\begin{thebibliography} {}

\bibitem{Ge_Ou}
Z.M. Ge and C.Y. Ou, Chaos, Solitons and Fractals {\bf34}, 262 (2007).

\bibitem{Li_Liao_Yu} C. Li, X. Liao and J. Yu, Phys. Rev. E
{\bf68}, 067203 (2003).

\bibitem{Li_Chen}
C. Li and G. Chen, Physica A {\bf341}, 55 (2004).

\bibitem{Li_Peng}
C. Li and G. Peng, Chaos, Solitons and Fractals {\bf22}, 443 (2004).

\bibitem{Zhang et al}
W. Zhang, S. Zhoua, H. Lib and H. Zhua, Chaos, Solitons and Fractals
{\bf42}, 1684 (2009).

\bibitem{Matignon}
D. Matignon, Stability results on fractional differential equations
with applications to control processing. In: Proc. of IMACS-SMC, pp.
963-968, Lille, France (1996).

\bibitem{Chen_Moore}
Y. Chen and K. Moore, Nonlinear Dyn. {\bf29}, 191 (2002).

\bibitem{Deng_Wu}
W. Deng, Y. Wu and C. Li, Int. J. Bif. Chaos {\bf16}, 465 (2006).

\bibitem{Deng_Li}
W. Deng, C. Li and J. L\"{u}, Nonlinear Dyn. {\bf48}, 409 (2007).

\bibitem{Deng1}
W. Deng and C. Li, J. Phys. Soc. Jpn. {\bf74}, 1645 (2005).

\bibitem{Deng2}
W. Deng and C. Li, Physica A {\bf353}, 61 (2005).

\bibitem{Deng3}
W. Deng, Phys. Rev. E {\bf75}, 056201 (2007).

\bibitem{Zhou}
T. Zhou and C. Li, Physica D {\bf212}, 111 (2005).

\bibitem{Li_Deng}
C. Li, W. Deng and D. Xu, Physica A {\bf360}, 171 (2006).

\bibitem{Barbosa1}
R.S. Barbosa, J.A.T Machado and I.M. Ferreira, Nonlinear Dyn.
{\bf38}, 305 (2004).

\bibitem{Barbosa2}
R.S. Barbosa, J.A.T Machado and A.M. Galhano, J. Vibration and
Control {\bf13}, 1407 (2007).

\bibitem{Pires}
E.J.S. Pires, J.A.T Machado and P.B. Oliveira, Signal Processing
{\bf83}, 2377 (2003).

\bibitem{Li_Sun}
C. Li, W. Sun and J. Kurths, Physica A {\bf361}, 24 (2006).

\bibitem{Wang_He}
X. Wang, Y. He and M. Wang, Nonlinear Analysis {\bf71}, 6126 (2009).

\bibitem{Tang et al}
Y. Tang, Z. Wang and J. Fang, Chaos {\bf19}, 013112 (2009).

\bibitem{SCH08}
E. Sch\"{o}ll and H. G. Schuster (ed.), \textit{Handbook of chaos control} (Wiley-VCH, Weinheim, 2008),
second completely revised and enlarged edition.

\bibitem{PYR92}
K. Pyragas, Phys. Lett. A \textbf{170}, 421 (1992).

\bibitem{PYR06}
K. Pyragas, Phil. Trans. R. Soc. A \textbf{364}, 2309 (2006).

\bibitem{PT93}
K. Pyragas and A. Tama\v{s}evi\v{c}ius, Phys. Lett. A \textbf{180}, 99 (1993).

\bibitem{BDG94} 
S. Bielawski, D. Derozier, and P. Glorieux, Phys. Rev. E \textbf{49}, R971 (1994).

\bibitem{PBA96} 
T. Pierre, G. Bonhomme, and A. Atipo, Phys. Rev. Lett. \textbf{76}, 2290 (1996).

\bibitem{HCT97} 
K. Hall, D. J. Christini, M. Tremblay, J. J. Collins,
L. Glass, and J. Billette, Phys. Rev. Lett. \textbf{78}, 4518 (1997).

\bibitem{SBG97}
D.W. Sukow, M. E. Bleich, D. J. Gauthier, and J. E. S.
Socolar, Chaos \textbf{7}, 560 (1997).

\bibitem{KF00}
J. M. Krodkiewski and J. S. Faragher, J. Sound Vib. \textbf{234},
591 (2000).

\bibitem{FSK02}
T. Fukuyama, H. Shirahama, and Y. Kawai, Phys. Plasmas
\textbf{9}, 4525 (2002).

\bibitem{RP04}
M. G. Rosenblum and A. S. Pikovsky, Phys. Rev. Lett. \textbf{92},
114102 (2004).

\bibitem{LBJ04}
C. von Loewenich, H. Benner, and W. Just, Phys. Rev. Lett. \textbf{93}, 174101 (2004).

\bibitem{PHT05}
O. V. Popovych, C. Hauptmann, and P. A. Tass, Phys. Rev.
Lett. \textbf{94}, 164102 (2005).

\bibitem{HS05}
P. H\"{o}vel and E. Sch\"{o}ll, Phys. Rev. E \textbf{72}, 046203 (2005).

\bibitem{YWH06}
S. Yanchuk, M. Wolfrum, P. H\"{o}vel and E. Sch\"{o}ll, Phys. Rev. E \textbf{74}, 026201 (2006).

\bibitem{DHS07}
T. Dahms, P. H\"{o}vel and E. Sch\"{o}ll, Phys. Rev. E \textbf{76}, 056201 (2007).

\bibitem{SSG94}
J. E. S. Socolar, D. W. Sukow and D. J. Gauthier, Phys. Rev E \textbf{50}, 3245 (1994)

\bibitem{PYR95}
K. Pyragas, Phys. Lett. A \textbf{206}, 323 (1995).

\bibitem{AP04}
A. Ahlborn and U. Parlitz, Phys. Rev. Lett. \textbf{93}, 264101 (2004).

\bibitem{AP05}
A. Ahlborn and U. Parlitz, Phys. Rev. E \textbf{72}, 016206 (2005).

\bibitem{SS97}
H. G. Schuster and M. B. Stemmler, Phys. Rev. E \textbf{56}, 6410 (1997).

\bibitem{P01}
K. Pyragas, Phys. Rev. Lett. \textbf{86}, 2265 (2001).

\bibitem{PPK02}
K. Pyragas, V. Pyragas, I. Z. Kiss and J. L. Hudson, Phys. Rev. Lett. \textbf{89}, 244103 (2002).

\bibitem{PPK04}
K. Pyragas, V. Pyragas, I. Z. Kiss and J. L. Hudson, Phys. Rev. E \textbf{70}, 026215 (2004).

\bibitem{JBO97}
W. Just, T. Bernard, M. Ostheimer, E. Reibold and H. Benner, Phys. Rev. Lett. \textbf{78}, 203 (1997).

\bibitem{NAK97}
H. Nakajima, Phys. Lett. A \textbf{232}, 207 (1997).

\bibitem{NU98}
H. Nakajima and Y. Ueda, Physica D \textbf{111}, 143 (1998).

\bibitem{FFG07}
B. Fiedler, V. Flunkert, M. Georgi, P. H\"{o}vel and E. Sch\"{o}ll, Phys. Rev. Lett \textbf{98}, 114101 (2007).

\bibitem{JFG07}
W. Just, B. Fiedler, M. Georgi, V. Flunkert, P. H\"{o}vel and E. Sch\"{o}ll, Phys. Rev. E \textbf{76}, 026210 (2007).

\bibitem{PS07}
C. M. Postlethwaite and M. Silber, Phys. Rev. E \textbf{76}, 056214 (2007).

\bibitem{GU08}
A. Gjurchinovski and V. Urumov, Europhys. Lett. \textbf{84}, 40013 (2008).

\bibitem{GU10}
A. Gjurchinovski and V. Urumov, Phys. Rev. E \textbf{81}, 016209 (2010).

\bibitem{Caputo}
M. Caputo, {\it Elasticita Dissipacione} (Zanichelli, Bologna,
1969).

\bibitem{Miller_Ross}
K. S. Miller and B. Ross, {\it An Introduction to the Fractional
Calculus and Fractional Differential Equations} (Wiley, New York,
1993).

\bibitem{Deng_Lu}
W. Deng and J. L\"{u}, Chaos {\bf16}, 043120 (2006).

\bibitem{Podlubny}
I. Podlubny, {\it Fractional Differential Equations} (Academic Press,
San Diego, 1999).

\bibitem{BJS05}
A. G. Balanov, N. B. Janson and E. Sch\"{o}ll, Phys. Rev. E {\bf 71},
016222 (2005).

\bibitem{Diethelm_Ford}
K. Diethelm, N. J. Ford and A. D. Freed, Nonlinear Dyn. {\bf29}, 3 (2002); 
K. Diethelm, N. J. Ford, A. D. Freed and Y. Luchko, Comput.
Methods Appl. Mech. Engrg. {\bf194}, 743 (2005).


\end{thebibliography}
\end{document}